# Seasonal and Trend Forecasting of Tourist Arrivals: An Adaptive Multiscale Ensemble Learning Approach

Shaolong Sun[a,*], Dan Bi[a], Ju-e Guo[a], Shouyang Wang[b]

[a]School of Management, Xi'an Jiaotong University, Xi'an 710049, China

[b]Academy of Mathematics and Systems Science, Chinese Academy of Sciences, Beijing 100190, China

*Corresponding author. School of Management, Xi'an Jiaotong University, Xi'an 710049, China.

Tel.: +86 159 1105 6725; Fax: +86 29 82665049.

E-mail address: sunshaolong@xjtu.edu.cn (S. L. Sun).

**Abstract:**

Accurate seasonal and trend forecasting of tourist arrivals is a very challenging task. Considering the importance of seasonal and trend forecasting of tourist arrivals, limited research has previously focused on these arrivals. In this study, a new adaptive multiscale ensemble (AME) learning approach incorporating variational mode decomposition (VMD) and least square support vector regression (LSSVR) is developed for short-, medium-, and long-term seasonal and trend forecasting of tourist arrivals. In the formulation of our developed AME learning approach, the original tourist arrival series are first decomposed into trend, seasonal and remaining volatility components. Then, ARIMA is used to forecast the trend component, SARIMA is used to forecast the seasonal component with a 12-month cycle, and LSSVR is used to forecast the remaining volatility components. Finally, the forecasting results of the three components are aggregated to generate an ensemble forecasting of tourist arrivals by the LSSVR-based nonlinear ensemble approach. Furthermore, a direct strategy is used to implement multistep-ahead forecasting. Taking two accuracy measures and the Diebold-Mariano test, the empirical results demonstrate that our proposed AME learning approach can achieve higher level and directional forecasting accuracy compared with other benchmarks used in this study, indicating that our proposed approach is a promising model for forecasting tourist arrivals with high seasonality and volatility.

# 1 Introduction

Tourism is a comprehensive industry and an important driving force for economic development. It makes a considerable contribution to the development of national economies in the following aspects: industrial structure, social employment and investment environment. Furthermore, accurate forecasts of tourism arrivals provide crucial information for tourism safety emergency work, especially during peak tourist times. Scholars' research on tourism arrival forecasts has been in full swing in recent years. Tourism arrival forecasting methods can be broadly divided into four categories: time series models, econometric models, artificial intelligence techniques and qualitative methods (Sun et al. 2019). Additionally, comparing the single model, the combination forecasts can considerably improve the performance of the forecast model, and no single forecasting method has been found to outperform all others in all situations (Shen et al. 2009). The performance of combination forecasts is associated with the performance consistency of the individual forecasts they include, and the inclusion of up to three individual forecasts is most likely to result in accurate combination forecasts (Shen et al. 2011).

Considering the importance of seasonal and trend forecasting of tourist arrivals, limited research has previously focused on these arrivals. The accurate seasonal and trend forecasting of tourist arrivals is a very challenging task, and seasonal demand variations represent a central theme not only in the academic literature on tourism but also in the domains of policymaking and practical tourism management (Koenig-Lewis and Bischoff 2005). Seasonality is one of the most important features of tourism demand and has important impacts on many aspects of the tourism industry. Modeling seasonal variation in international tourism demand has become an important issue in tourism forecasting in recent years, and accurate forecasts of seasonal tourism demand are crucial for the formulation of effective marketing strategies and tourism policies for both the private and public sectors (Chen et al. 2017). Multiple kinds of research have attempted to use time series models and machine learning methods (usually one or two kinds), and there have been few attempts to combine both of them according to decomposed data features, such as trend, volatility and seasonal components.

In this study, a new adaptive multiscale ensemble (AME) learning approach incorporating variational mode decomposition (VMD) and least square support vector regression (LSSVR) is developed for short-, medium-, and long-term seasonal and trend forecasting of tourist arrivals. In the formulation of our developed AME learning approach, the original tourist arrival series are first decomposed into the trend, seasonal and remaining volatility components. Then, ARIMA is used to forecast the trend component, SARIMA is used to forecast the seasonal component with a 12-month cycle, and LSSVR is used to forecast the remaining volatility components. Finally, the forecasting results of the three components are aggregated to generate an ensemble forecast of tourist arrivals by the LSSVR-based nonlinear ensemble approach. Furthermore, a direct strategy is used to implement multistep-ahead forecasting.

Our research contributes to the tourism forecasting literature by proposing a decomposition-forecasting-integration ensemble learning approach based on the features of tourist arrival time series data. First, this article makes a detailed comparison and summary of the methodologies used in tourism demand forecast articles in the past ten years. Second, the decomposition-forecasting-integration ensemble learning method, which does not rely on other variables and greatly improves the forecasting performance, is applied to the work of tourism forecasting. Third, we decomposed original tourist arrivals into trend, period, and volatility using VMD, which is a new adaptive signal processing method that has a good processing effect on nonstationary and nonlinear signals. Fourth, the AME method proposed in this article performs short-, medium-, and long-term forecasting of tourist arrivals and is empirically proven to have superior forecasting performance, which fully illustrates our AME learning approach's performance in forecasting for different durations of applicability.

## 2 Literature review

After decades of research on the tourism forecast literature, three main categories of quantitative forecasting methods have been identified: time series models, econometric approaches, and artificial intelligence (AI) models (Peng et al. 2014). Time series analysis is based on the time series data obtained by system observation. It is a method for establishing mathematical models through curve fitting and parameter estimation. The advantage is that only a series of time series data is required because it cannot describe the impact factors. Econometric tools require considerable time and work to explain different impact factors, and the combination of econometric approaches with time series models has prevailed in recent studies (Cao et al. 2017; Fildes et al. 2011).

The adaptive ensemble learning approach we propose is a comprehensive methodology based on the idea of combination forecasts. The main idea is to complete learning tasks by constructing and combining multiple learners. Ensemble learning often achieves significantly better generalization performance than a single learner. Therefore, combining forecasts based on different methods or data has emerged as one of the most important methods for improving forecasting performances. The combination of quantitative and judgmental forecasts adds a promising dimension to forecasting and has been a key research area over the past three decades (Song et al. 2013). The combination forecasts model is more accurate in tourism forecasts than the single model. Some research results suggest that, overall, combination forecasts can improve forecast model performance, as they are superior to the best individual forecasts (Coshall and Charlesworth 2011; Shen et al. 2008); minimally, all the combined forecasts are not outperformed by the worst single model forecasts (Wong et al. 2007). Although combination forecast models have a long-established position in forecasting, nonlinear combination models are comparatively rare. Some studies consider a set of nonlinear combination forecast models to have better forecasting accuracy performance(Cang 2014; Wong et al. 2006). Some researchers have noted the volatility of forecast tourism demand and used volatility models to analyze the impact of negative shocks (Coshall 2009).

Seasonal demand variations represent an important status in both academic researchers and policymakers (Koenig-Lewis and Bischoff 2005), and there are a variety of models describing seasonal factors. Before the rise of big data, econometric methods and time series models were widely used to forecast seasonal tourism trends based on monthly or quarterly time series data (Chu 2008; Gil-Alana et al. 2008; Liu et al. 2018; Song and Witt 2006), and empirical studies provide a comparison of the performance of commonly used econometric and time series models in forecasting seasonal tourism demand (Shen et al. 2009). The empirical results suggest that no single forecasting technique is superior to others in all situations.

Econometrics are also used in tourism forecasting along with time series models (Gil-Alana et al. 2008; Song et al. 2003). However, because of the more difficult model factors of econometrics and the time cost, time series models are currently the more commonly used models in tourism forecasting, and time series models also show superior performance in dealing with seasonality. Several works have used time series models and econometrics to forecast tourism, and most of the forecast targets are tourist arrivals (du Preez and Witt 2003; Song et al. 2003). Some papers used aggregated machine learning models to overcome the difficulty of forecasting and improve forecasting performance (Palmer et al. 2006; Shen et al. 2008; Wu et al. 2012). From these studies, it can be seen that in the comparison of seasonal tourism demand forecasting, econometric models and time series models have advantages. Econometric models are good at explaining the effects of economic variables, while time series models have better fitting performance.

Artificial intelligence models can also capture seasonal trends in tourism well; hence, in recent years, an increasing number of scholars have applied artificial intelligence (AI) technology to the field of tourism forecasts. AI forecasting models, including neural networks, rough sets theory, fuzzy time series theory, gray theory, genetic algorithms, and expert systems, tend to perform better than traditional forecasting methods (Peng et al. 2014). Research integrates AI, such as rough set theory, to capture useful information from a set of raw hybrid data and discover knowledge from the data in the form of decision rules (Goh and Law 2003). Additionally, SVR with GA has been used in tourism demand forecasting compared with Back Propagation Neural Network(BPNN) and ARIMA (Chen and Wang 2007), as well as the use of gray theory and fuzzy time series, which do not need large samples and long past time series to estimate tourist arrivals (Wang 2004). Combining AI models to obtain accurate forecasting results is becoming prevalent in recent studies.

Time series models are also combined with artificial intelligence (AI) models to predict the number of tourist arrivals, such as nonlinear dynamics in a time series of airport arrivals, which proves that the reconstruction approach offers better results in sign prediction and the learning approach in point prediction (Olmedo 2016). Some researchers have used a hybrid intelligent method to combine original time series data with AI models, which also presents superior performance (Kim et al. 2010; Li et al. 2018; Shahrabi et al. 2013). Some scholars compared time series models with AI models in tourism forecasting and found that no single model can provide the best forecasts for any country in the short, medium and long run (Hassani et al. 2017). This shows that the integrated model can combine the advantages of the time series model with the AI model, which can greatly improve the work of tourism forecasting. Some scholars combined the characteristics of big data, added the Internet search index to the variables that affect the effect of tourism forecasting, combined AI models and time series models to make travel forecasts, and obtained better forecasting results (Gunter and Önder 2016; Pan et al. 2012; Pan and Yang 2016; Yang et al. 2015). However, at present, most of the research combines AI and time series, but none of the original time series data decomposition and analysis integrated approaches, and therefore does not work well in the applicability of their tools.

To further illustrate the methodology of tourism forecasting in recent years, specific details and applicable tools are shown in Table 1. As shown in the table, the forecasting body, methodology, and parameters for evaluating the forecast performance of each article are listed in detail. It seems that the arrival of tourists is an important indicator of tourism demand. In recent years, a large number of articles using artificial intelligence methods have emerged, as well as studies combining time series models with AI methodologies. In artificial intelligence research, many articles apply neural network methods. The data frequency is mainly monthly or quarterly.

Overall, there were two main limitations in the previous studies related to forecasting tourism demand with AI and time series models. First, most literature only uses one or two AI or time series models, the applicability of each model is limited, and conflicting conclusions still exist regarding which models generate the most accurate forecasts under different conditions. Each method has its advantages in dealing with a particular problem, but none is universally superior (Peng et al. 2014). Second, most studies do not consider the data features, such as the trend and the seasonality of the data, resulting in AI or time series models that do not work well in the applicability of their tools. To compensate for the literature gaps shown above, a new adaptive multiscale ensemble (AME) learning approach incorporating variational mode decomposition (VMD) and least square support vector regression (LSSVR) is developed for short-, medium-, and long-term seasonal and trend forecasting of tourist arrivals.

**Table 1** An overview of selected tourism forecasting studies.

| References | Region focused | Research objects | Data frequency | Methodologies | Performance measure | Variables |
|---|---|---|---|---|---|---|
| (Rivera 2016) | Puerto Rico | Hotel nonresident registrations | Monthly | DLM | MAE, MAPE, RMSE | Google trend data, NHNR |
| (Coshall and Charlesworth 2011) | UK | Outbound tourism | Quarterly | Combination forecasts, goal programming | MAPE | Air passengers |
| (Peng et al. 2014) | Europe | Tourism demand | Monthly, quarterly | Meta-analysis | MAPE, RMSE | Tourist arrivals |
| (Chu 2011) | Macau | Tourism demand | Monthly | Piecewise linear method | MAPE, RMSE | Tourist arrivals |
| (Shen et al. 2008) | USA | UK outbound leisure tourism demand | Quarterly | simple average combination, variance-covariance combination, discounted MSFE method | MAPE | Tourist arrivals, GDP, relative tourism price |
| (Chu 2008) | Asian-Pacific | Tourism demand | Monthly, quarterly | ARAR model | MAPE, RMSE | Tourist arrivals |
| (Athanasopoulos et al. 2017) | Australia | Tourism demand | Quarterly | Bagging ensemble, model selection | MAPE, RMSE | Tourist arrivals, economic and dummy variables |
| (Wong et al. 2006) | Hong Kong | Tourism demand | Annual | Bayesian vector autoregressive | MAE, RMSE | Tourist arrivals, economic and dummy variables |
| (Kim et al. 2010a) | Hong Kong | Tourism demand | Monthly | Bias-corrected | Prediction | Tourist arrivals, |
| | | | | bootstrap, AR models | interval | dummy variables |
| (Bangwayo-Skeete and Skeete 2015) | Caribbean | Tourism demand | Monthly | AR-MIDAS | MAPE, RMSE, DM | Tourist arrivals, Google trend data |
| (Shen et al. 2011) | UK | Outbound tourism demand | Quarterly | Combination forecasts and single forecasts | MAPE | Tourist arrivals, economic and dummy variables |
| (Andrawis et al. 2011) | Egypt | Inbound tourism | Monthly | Combination forecasts | MAPE, MAE, Wilcoxon test | NN3, M3, tourist arrivals |
| (Song et al. 2013) | Hong Kong | Tourism demand | Quarterly, annual | ADLM, judgmental forecasting | MAPE, RMSE | Tourist arrivals, tourist expenditure, and economic variables |
| (Coshall 2009) | UK | Outbound Tourism | Quarterly | GARCH, ES, combining forecast | MAPE, RMSE, encompassing tests | Tourist arrivals |
| (Olmedo 2016) | Spain | Airport arrivals | Daily | ANN, PSR | NMSE, DC | Air arrivals |
| (Palmer et al. 2006) | Spain | Tourism demand | Quarterly | MLP | RMSE, MAPE, U-Theil | Tourism expenditure |
| (Shahrabi et al. 2013) | Japan | Inbound tourism | Monthly | MGFFS | RMSE, MAPE | Tourist arrivals |
| (Shen et al. 2009) | UK | Outbound tourism | Quarterly | SARIMA, BSM, TVP | RMSE, MAPE | Tourist arrivals, economic variable |
| (Fildes et al. 2011) | UK | Air travel demand | Annual | ADLM, TVP, VAR | MAE, RMSE | Air passengers and |
| | | | | | | economic variable |
| (Hassani et al. 2017) | European countries | Tourism demand | Monthly | RSSA, NN, MA, ARIMA, ARFIMA, ES | RMSE, DC | Tourist arrivals |
| (Gunter and Önder 2016) | Vienna | Tourism demand | Monthly | Bayesian FAVAR | RMSE, MAE | Tourist arrivals and Google trend data |
| (Pan and Yang 2016) | Charleston county | Hotel demand | Weekly | ARIMAX | MAPE, RMSE | Hotel occupancy, search engine queries, website traffic, weather information |
| (Hirashima et al. 2017) | Hawaii | Tourism demand | Monthly, quarterly | MIDAS | MAPE, RMSE, DM | Tourist arrivals and economic variables |
| (Gunter and Önder 2015) | Paris | Inbound tourism | Monthly | EC-ADLM, VAR, Bayesian VAR, TVP, ARMA, ETS | RMSE, MAE | Tourist arrivals in hotels and economic variable |
| (Song and Witt 2006) | Macau | Inbound tourism | Quarterly | VAR | AIC, LL, SBC | Tourist arrivals and economic variables |
| (Chu 2004) | Singapore | Inbound tourism | Monthly | Cubic polynomial model | MAPE | Tourist arrivals |
| (Chen et al. 2012) | Taiwan | Inbound tourism | Monthly | EMD, BPNN | MAD, MAPE, RMSE | Tourist arrivals |
| (Chu 2009) | Asian-Pacific region | Tourism demand | Monthly, quarterly | ARIMA, ARAR, ARFIMA | ME, MPE, MAPE, RMSE | Tourist arrivals |
| (Li et al. 2017) | Beijing | Tourism demand | Monthly | GDFM, PCA | MAE, MAPE | Tourist arrivals and Baidu index |
| (Song et al. 2011) | Hong Kong | Inbound tourism | Monthly | STSM, TVP | MAPE, RMSE | Tourist arrivals and economic variables |
| (Goh and Law 2003) | Hong Kong | Inbound tourism | Quarterly | Rough set theory | Accuracy of classification | Tourist arrivals and economic variables |
| (Athanasopoulos and Hyndman 2008) | Australian | Inbound tourism | Quarterly | SSME, ES | RMSE, ME, MAE, MAPE | Tourist arrivals and economic variables |
| (Song et al. 2003) | Hong Kong | Inbound tourism | Quarterly | ADLM | MSE, $R^2$ | Tourist arrivals and economic variables |
| (Wang 2004) | Taiwan | Inbound tourism | Annual | NN, fuzzy theory, GM(1,1) model | RPE | Tourist arrivals |
| (Wu and Cao 2016) | China | Inbound tourism | Monthly | Seasonal index adjustment, SVR, FOA | RMSE, MAPE, $R$ | Tourist arrivals |
| (Chen and Wang 2007) | China | Tourism demand | Quarterly | SVR, GA | NMSE, MAPE | Tourist arrivals |
| (Athanasopoulos et al. 2011) | Australia, Hong Kong, New | Tourism demand | Monthly, quarterly, annual | ARIMA, ES, ADLM, TVP, VAR | MAPE, MASE, MdASE, | Tourist arrivals and economic variables |
| | Zealand | | | | | |
| (Guizzardi and Mazzocchi 2010) | Italian | Hotels demand | Quarterly | STS, LCC, XCV | MAPE, Wilcoxon signed-rank test | Hotel overnight stays and economic variable |
| (Song et al. 2010) | Hong Kong | Inbound tourism | Monthly | ADLM | MAPE, RMSE | Tourist arrivals and expenditure, economic variable |
| (Chan et al. 2010) | Hong Kong | Inbound tourism | Quarterly | CUSUM, combination forecast, quadratic programming | MAPE, RMSE | Tourist arrivals and economic variables |
| (Song et al. 2003) | Denmark | Inbound tourism | Annual | ADLM, ECM, TVP, VAR, ARIMA | MAPE, RMSE | Tourist arrivals and economic variables |
| (Wong et al. 2007) | Hong Kong | Inbound tourism | Quarterly | ARIMA, ADLM, ECM, VAR, combining forecast | MAPE, RMSE, MAE | Tourist arrivals and economic variables |
| (Gil-Alana et al. 2008) | Canary island | Tourism demand | Monthly | SARFIMA | MAPE, MSE, RMSE, MAD | Tourist arrivals |
| (du Preez and Witt 2003) | Seychelles | Inbound tourism | Monthly | ARIMA, SSM | MAE, RMSE, MAPE | Tourist arrivals and economic variables |
| (Sun et al. 2016) | China | Inbound tourism | Annual | MCGM(1,1), CSO | MAPE, MSE | Tourist arrivals |

**Notes:** The number of hotel nonresident registrations (NHNR); dynamic linear model (DLM); autoregressive mixed-data sampling (AR-MIDAS) models;

autoregressive distributed lag model (ADLM); phase-space reconstruction (PSR); modular genetic-fuzzy forecasting system (MGFFS); structural time series model (BSM); recurrent singular spectrum analysis (RSSA) model; factor-augmented vector autoregression (FAVAR); Akaike information criterion (AIC); log likelihood value (LL); Schwarz Bayesian criterion (SBC); mean absolute deviation (MAD); self-exciting threshold autoregression models (SETAR); mean error (ME); mean percentage error (MPE); generalized dynamic factor model (GDFM); principal component analysis (PCA); structural time series model (STSM); state space models with exogenous variables (SSME) model; global vector autoregressive (GVAR) model; relative percentage error (RPE); fruit fly optimization algorithm (FOA); genetic algorithm (GA); chaotic genetic algorithm (CGA); mean absolute scaled error (MASE); median absolute scaled error (MdASE); structural time series (STS); latent cyclical component model (LCC); specific economic explanatory variables (XCV); fuzzy c-means (FCM); logarithm least-squares support vector regression (LLS-SVR); error correction model (ECM); state space model (SSM); Markov-chain gray model (MCGM); cuckoo search optimization algorithm (CSO);

## 3 Related methodology

Before presenting our proposed adaptive multiscale ensemble learning approach, we introduce some methods used in this approach.

### *3.1 Variational mode decomposition*

Variational mode decomposition (VMD) is an effective signal processing technique proposed by Dragomiretskiy (Dragomiretskiy and Zosso 2014). VMD has been widely used in practical applications. Previous literature has shown the beneficial power of VMD in the signal denoising field against other signal decomposition algorithms, such as wavelet transform (WT) and empirical mode decomposition (EMD). The main purpose of VMD is to divide the original signal $y$ into $k$ discrete band-limited modes, where each mode is required to compact around a center pulsation $\omega_k$ determined along with the decomposition process. The bandwidth of each mode $y_k$ can be obtained by the following procedures: (1) to obtain a unilateral frequency spectrum, we compute the associated analytic signal for each mode $y_k$ using the Hilbert transform; (2) to shift the mode's frequency spectrum to baseband, we combine with an exponential tuned to the respective estimated center frequency; (3) the bandwidth of each mode $y_k$ can be estimated by the Gaussian smoothness of the demodulated signal. Then, the constrained variational problem is provided as follows:

$$\min_{\mu_k,\omega_k} \left\{ \sum_{k=1}^{K} \left\| \partial_t \left[ \left( \delta(t) + \frac{j}{\pi t} \right) * y_k(t) \right] e^{-j\omega_k t} \right\|_2^2 \right\} \tag{1}$$

subject to

$$\sum_{k=1}^{K} y_k = y \tag{2}$$

where $y$ is the original signal, $K$ is the number of modes, $\delta$ is the Dirac distribution, $t$ is the time script, $*$ denotes the convolution, and $\{y_k\} = \{y_1, y_2, \ldots, y_K\}$ and $\{\omega_k\} = \{\omega_1, \omega_2, \ldots, \omega_K\}$ represent the set of all modes and their center pulsations.

To address the constrained variational problem above, both quadratic penalty terms and Lagrangian multipliers are introduced. Then, the augmented Lagrangian can be represented as follows:

$$\ell(y_k, \omega_k, \lambda) = \alpha \sum_{k=1}^{K} \left\| \partial_t \left[ \left( \delta(t) + \frac{j}{\pi t} \right) * y_k(t) \right] e^{-j\omega_k t} \right\|_2^2 + \left\| y(t) - \sum_{k=1}^{K} y_k(t) \right\|_2^2 + \left\langle \lambda(t), y(t) - \sum_{K=1}^{k} y_k(t) \right\rangle \tag{3}$$

where $\alpha$ is the balancing parameter of the data-fidelity constraint, $\lambda$ denotes the Lagrange multipliers for tightening restraint, and $\left\| y(t) - \sum_{k=1}^{K} y_k(t) \right\|_2^2$ represents a quadratic penalty term for accelerating the rate of convergence.

Furthermore, the solution of Eq. (3) is found in a sequence of iterative suboptimizations called the alternate direction method of multipliers (ADMM). Consequently, the solutions for $y_k$ and $\omega_k$ can be obtained as follows:

$$\hat{y}_k^{n+1}(\omega) = \frac{\hat{y}(\omega) - \sum_{i \neq k} \hat{y}_i(\omega) + \frac{\hat{\lambda}(\omega)}{2}}{1 + 2\alpha(\omega - \omega_k)^2} \tag{4}$$

$$\omega_k^{n+1} = \frac{\int_0^\infty \omega |\hat{y}_k(\omega)|^2 \, d\omega}{\int_0^\infty |\hat{y}_k(\omega)|^2 \, d\omega} \tag{5}$$

where $\hat{y}_k^{n+1}(\omega)$, $\hat{\lambda}(\omega)$, $\hat{y}_i(\omega)$ and $\hat{y}(\omega)$ represent the Fourier transforms of $y_k^{n+1}(t)$, $\lambda(t)$, $y_i(t)$ and $y(t)$, respectively, and $n$ is the number of iterations. For further details on the VMD method, please refer to Dragomiretskiy and Zosso (2014).

## *3.2 Autoregressive integrated moving average*

The autoregressive integrated moving average (ARIMA) developed by Box and Jenkins (1970) is the most widely used time series model. The process to build this model was designed to take advantage of associations in the sequentially lagged relationships that usually exist in data collected periodically. The general form of the ARIMA mode is as follows:

$$\phi(B)\delta(B)x_t = \theta(B)\varepsilon_t \qquad (6)$$

where $x_t$ is the actual value at time $t$; $\varepsilon_t$ is the error term and $\varepsilon_t \sim iid(0,\sigma^2)$; $B$ is the backshift operator defined by $B^a x_t = x_{t-a}$; $\phi(B)$, $\theta(B)$ and $\delta(B)$ represent polynomials of $B$, $\phi(B)=1-\phi_1 B-\cdots-\phi_p B^p$, $\theta(B)=1-\theta_1 B-\cdots-\theta_q B^q$ and $\delta(B)=(1-B)^d$. Therefore, the model is summarized as ARIMA (p,d,q), and the three parameters $p$, $d$ and $q$ in ARIMA (p,d,q) are determined by the model information criterion, such as the Akaike information criterion (AIC).

## *Seasonal autoregressive integrated moving average*

A time series $\{X_t\}$ is a seasonal $ARIMA(p,d,q)(P,D,Q)_S$ process with a period $S$ if $d$ and $D$ are nonnegative integers and if the differenced series $Y_t = (1-B)^d (1-B^S)^D X_t$ is a stationary autoregressive moving average (ARMA) process defined by the expression as follows:

$$\phi(B)\Phi(B^S)Y_t = \theta(B)\Theta(B^S)\varepsilon_t \qquad (7)$$

where $B$ is the backshift operator defined by $B^a X_t = X_{t-a}$; $\phi(z)=1-\phi_1 z-\cdots-\phi_p z^p$, $\Phi(z)=1-\Phi_1 z-\cdots-\Phi_Q z^Q$; $\theta(z)=1-\theta_1 z-\cdots-\theta_q z^q$, $\Phi(z)=1-\Phi_1 z-\cdots-\Phi_Q z^Q$; $\varepsilon_t$ is identically and normally distributed with mean zero, variance $\sigma^2$; and $\text{cov}(\varepsilon_t, \varepsilon_{t-k}) = 0, \quad \forall k \neq 0$, that is, $\{\varepsilon_t\} \sim WN(0, \sigma^2)$.

The parameters $p$ and $P$ represent the nonseasonal and seasonal autoregressive polynomial order, respectively, and the parameters $q$ and $Q$ represent the nonseasonal and seasonal moving average polynomial order, respectively. As discussed above, the parameter $d$ represents the order of normal differencing, and the parameter $D$ represents the order of seasonal differencing. From a practical perspective, fitted seasonal ARIMA models provide linear state transition equations that can be applied recursively to produce single and multiple interval forecasts. Furthermore, seasonal ARIMA models can be readily expressed in state space form, thereby allowing adaptive Kalman filtering techniques to be employed to provide a self-tuning forecast model.

### *3.3 Multilayer perceptron neural networks*

Artificial neural networks (ANNs) have become a widely used technique for exploring the dynamics of a variety of financial applications. Since foreign exchange markets are highly volatile, nonlinear and irregular, several neural networks have been applied to forecast foreign exchange rates, such as a multilayer perceptron (MLP), a radial basis function neural network (RBFNN) and a recurrent neural network (RNN).

The multilayer perceptron (MLP) neural network makes a complex mapping from inputs onto appropriate outputs and thus enables the network to approximate almost any nonlinear function even with one hidden layer. The relationship between the input variables ($y_{t-1}, y_{t-2}, \cdots, y_{t-p}$) and the output variable ($y_t$) has the following form:

$$y_t = \alpha_0 + \sum_{j=1}^{q} \alpha_j f\left(\beta_{oj} + \sum_{i=1}^{p} \beta_{ij} y_{t-i}\right) + \varepsilon_t \quad (8)$$

where $\alpha_j\left(j = 0,1,\cdots,q\right)$ and $\beta_{ij}\left(i = 0,1,\cdots,p; j = 1,2,\cdots,q\right)$ are the network parameters and $p$ and $q$ are the numbers of input nodes and hidden nodes, respectively. The activation function of the hidden layer often uses the logistic function $f\left(y\right) = 1/\left(1+\exp\left(-y\right)\right)$.

Backpropagation (BP) algorithms are one of the most commonly used training algorithms for MLP networks that minimize the total square errors of in-sample forecasting results. One challenge is to determine the number of neurons in each layer, the number of hidden layers, momentum parameters, and learning rates. To explore the optimal architecture of MLP networks, these parameters can be determined utilizing trial and error or particle swarm optimization algorithms. MLP networks can be utilized for foreign exchange rate modeling and forecasting. However, it is difficult to identify the optimal input size. The underlying economic theory is used to help determine the optimal input size. In this study, we use the autoregressive model to identify the input size.

### *3.4 Least square support vector machine*

The support vector machine (SVM) originally proposed by Cortes and Vapnik (1995) is based on statistical learning theory and the principle of structural risk minimization, which possesses good performance even for small samples. However, it is time consuming and leads to high computational costs when dealing with a large-scale problem. Hence, Suykens and Vandewalle (1999) proposed the least square support vector regression (LSSVR).

The basic idea of support vector regression (SVR) is to map the original data into a high-dimensional feature space and perform a linear regression in the space. It can be expressed as follows:

$$f(x) = w^T \varphi(x) + b \tag{9}$$

where $\varphi(x)$ denotes a nonlinear mapping function, $f(x)$ represents the estimation value, and $w^T$ and $b$ are the weights and basis, respectively.

It can be transformed into the following optimization problem:

$$\begin{aligned} &\min \quad \frac{1}{2} w^T w + C \sum_{t=1}^{T} \left( \xi_t + \xi_t^* \right) \\ &s.t. \quad \begin{cases} w^T \varphi(x_t) + b - y_t \le \varepsilon + \xi_t \\ y_t - \left( w^T \varphi(x_t) + b \right) \le \varepsilon + \xi_t^* \\ \xi_t, \xi_t^* \ge 0, \quad (t = 1, 2, \cdots, T) \end{cases} \end{aligned} \tag{10}$$

where $C$ denotes the penalty parameter and $\xi_t$ and $\xi_t^*$ represent the nonnegative slack variables.

It is time consuming to solve the above problem, and LSSVR is proposed to transform the problem as follows:

$$\begin{aligned} &\min \quad \frac{1}{2} w^T w + C \sum_{t=1}^{T} e_t^2 \\ &s.t. \quad y_t = w^T \varphi(x_t) + b + e_t, (t = 1, 2, \cdots, T) \end{aligned} \tag{11}$$

where $e_t$ denotes the slack variable. Generally, the parameters of SVR and LSSVR have a considerable influence on the accuracy of the regression estimation. Therefore, 5-fold cross-validation is employed to automatically select the optimal parameters of SVR and LSSVR in this study.

# 4 The framework of the adaptive multiscale ensemble learning approach

In this study, the h-step-ahead forecasting horizons are used to evaluate the forecasting performance of our proposed adaptive multiscale ensemble learning approach. Given a time series $y_t, (t = 1, 2, \cdots, n)$, we design *h*-step-ahead forecasting for $\hat{y}_{t+h}$:

$$\hat{y}_{t+h} = f\left(y_t, y_{t-1}, \ldots, y_{t-(l-1)}\right) \tag{12}$$

where $\hat{y}_{t+h}$ is the *h*-step-ahead forecasted value at time $t$, $y_t$ is the actual value at time $t$, and $l$ denotes the lag orders that are selected by autocorrelation and partial correlation analysis.

The framework of our proposed adaptive multiscale ensemble learning approach is shown in **Fig. 1**. Our proposed adaptive multiscale ensemble learning approach mainly consists of four steps as follows:

**Step 1**. A mode number fluctuation algorithm is used to determine the number $k$ of modes.

**Step 2**. The original tourist arrival time series $\{y_1, y_2, \ldots, y_n\}$ is decomposed into $k$ modes via the VMD method.

**Step 3**. Each mode series is forecasted by using ARIMA, SARIMA, and LSSVR.

**Step 4**. The forecasting results of all extracted modes are fused to generate a final forecasting result $\hat{y}_t$, using LSSVR as an ensemble learning method.

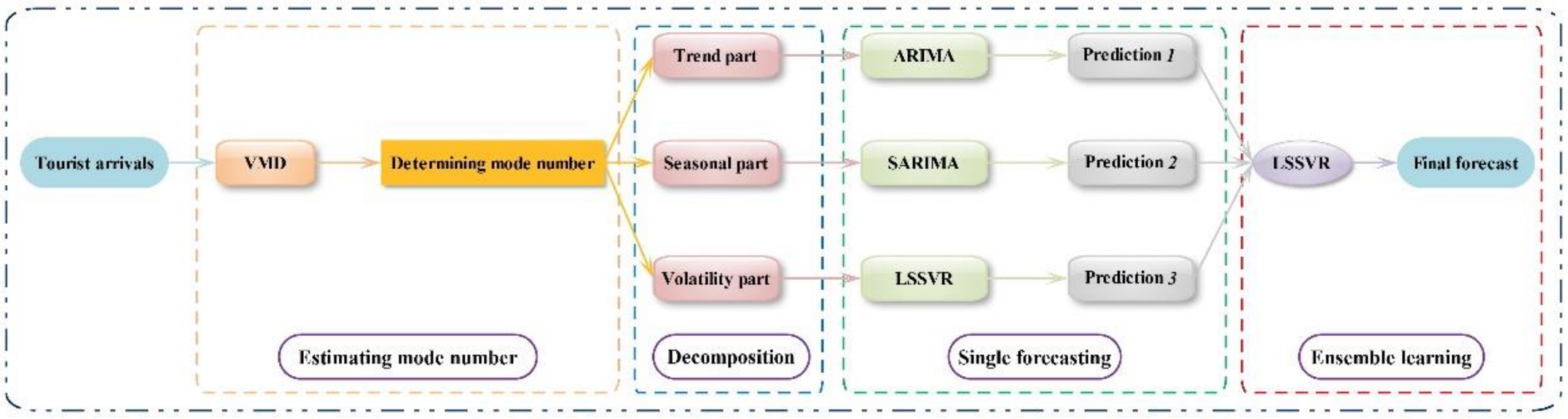

**Fig. 1** The framework of our proposed adaptive multiscale ensemble learning approach.

## 5 Empirical study

In this section, there are two main problems: (1) to evaluate the performance of our proposed adaptive multiscale ensemble learning approach for tourist arrival forecasting and (2) to demonstrate the superiority of our proposed adaptive multiscale ensemble learning approach in comparison with several other benchmarks. To accomplish these two tasks, we use tourist arrival data from Beijing city and Hainan province to check the forecasting performance of our proposed adaptive multiscale ensemble learning approach. The research data and evaluation criteria are introduced in **Section 5.1**, empirical results are presented in **Section 5.2**, and further discussions are given in **Section 5.3**.

### *5.1 Data descriptions and evaluation criteria*

In this study, the monthly tourist arrivals of Beijing city and Hainan province were collected from the Wind Database (http://www.wind.com.cn/). The data of Beijing city's tourist arrivals range from January 2010 to October 2019 with 118 observations, and the data of Hainan province's tourist arrivals cover from January 2003 to October 2019 with 203 observations. The dataset is divided into in-sample (training sample) and out-of-sample (testing sample), as provided in **Table 2**. The detailed data are not listed here but can be accessed from the wind database or obtained from the authors.

**Table 2** In-sample and out-of-sample datasets of the tourist arrival data.

| Sites | Sample type | From | To | Sample size |
|---|---|---|---|---|
| Beijing | total sample | January 2010 | October 2019 | 118 |
| | in-sample | January 2010 | October 2017 | 94 |
| | out-of-sample | November 2017 | October 2019 | 24 |
| | total sample | January, 2003 | October 2019 | 203 |
| Hainan | in-sample | January, 2003 | June, 2016 | 162 |
| | out-of-sample | July 2016 | October 2019 | 40 |

**Table 3** provides the descriptive statistics of the two tourist arrival data. We see the difference in the statistical features among the subsets. Specifically, these analyses show that the fluctuation of tourist arrival time series is unstable. Skewness analysis was adopted to depict the symmetry of the subset; the greater the absolute skewness value is, the more obvious the asymmetry. Additionally, kurtosis was measured to depict the steepness of the subset. For kurtosis, values greater than 0 indicate that the distribution of the dataset is steeper than the standard Gaussian distribution; in contrast, values less than 0 indicate that the distribution of the subset series is less steep than the standard Gaussian distribution; moreover, if the value is equal to 0, then it shows that variables have the same distribution as the standard Gaussian distribution.

**Table 3** Descriptive statistics of in-sample tourist arrival data.

| Exchange rate | Minimum | Maximum | Mean | Std.$^a$ | Skewness | Kurtosis |
|---|---|---|---|---|---|---|
| Beijing | 423.00 | 3553.20 | 1859.74 | 806.53 | 0.27 | 1.99 |
| Hainan | 34.42 | 425.24 | 177.16 | 77.82 | 0.82 | 3.04 |

Note Std.* refers to the standard deviation.

To evaluate the forecasting performance of our proposed adaptive multiscale ensemble learning approach from different perspectives, such as level forecasting and directional forecasting, two main evaluation criteria, i.e., mean absolute percentage error (MAPE) and directional symmetry (DS), are chosen as follows:

$$MAPE = \frac{1}{T}\sum_{t=1}^{T}\left|\frac{y_t - \hat{y}_t}{y_t}\right| \times 100\% \tag{13}$$

$$DS = \frac{1}{T}\sum_{t=1}^{T} d_t \times 100\%,\ \ d_t = \begin{cases} 1 & if \quad (y_{t+1} - y_t)(\hat{y}_{t+1} - y_t) \geq 0 \\ 0 & otherwise \end{cases} \tag{14}$$

where $\hat{y}_t$ and $y_t$ denote the forecast value and the actual value, respectively, and $T$ represents the number of sample observations.

However, to evaluate the forecasting performance of our proposed adaptive multiscale ensemble learning approach from a statistical perspective, two statistical tests, Diebold-Mariano (DM) (Diebold and Mariano 2002; Sun et al. 2017) and Pesaran-Timmermann (PT) (Pesaran and Timmermann 1992; Sun et al. 2017), are performed. The DM test aims to check the null hypothesis of equality of expected forecast accuracy against the alternative of different forecasting abilities across models. In this study, the mean square error (MSE) is considered as the DM loss function, and our proposed adaptive multiscale ensemble learning approach is compared against the other benchmark models under study. The PT test is used to examine whether the directional changes of the actual and forecasted values correspond with one another. In other words, it determines how well increases and decreases in the predicted value follow the real increases and decreases of the original time series. The null hypothesis is that the model under study has no power in forecasting foreign exchange rates. For a detailed mathematical derivation of the DM and PT statistical tests, please refer to Diebold and Mariano (2002) and Pesaran and

Timmermann (1992).

## *5.2 Empirical results*

To verify the superiority of our proposed adaptive multiscale ensemble learning approach, there are eight forecasting models built and used as benchmarks (i.e., five single models including the autoregressive integrated moving averaging (ARIMA) model, seasonal ARIMA (SARIMA) model, multilayer perceptron (MLP) neural network, support vector regression (SVR) and least square SVR (LSSVR), and three decomposition ensemble learning approaches including VMD-MLP, VMD-SVR, and VMD-LSSVR. The reasons for choosing these benchmarks are as follows: (1) ARIMA is a very popular benchmark model employed by Sun et al. (2017). (2) SARIMA has a noticeable impact on tourist arrival forecasting as one of the periodical and seasonal models presented in the econometrics literature (Brooks, 2014) and has shown its superiority in modeling tourist arrivals. (3) MLP, SVR and LSSVR are the most widely used machine learning techniques in tourist arrival forecasting, as introduced in **Section 2**. (4) The VMD-MLP, VMD-SVR and VMD-LSSVR decomposition ensemble approaches verify the capability of adaptive multiscale ensemble learning in our proposed approach.

The parameters of the ARIMA and SARIMA models are estimated by employing an automatic model selection algorithm implemented using the “forecast” program package in R software. For machine learning-based techniques, the MLP network employs standard two-layer neural network structures, including a hidden layer and an output layer. The number of hidden nodes is set to 20, as Godarzi (Godarzi et al.

2014) notes that a small number of hidden neurons results in the inaccuracy of the correlation between inputs and outputs, while too large a number of hidden neurons results in local optimums. The typical number of hidden neurons is in the range of 5 to 100, and it is unnecessary to employ cross-validation. The logistic sigmoid function is selected as the activation function, and the backpropagation algorithm is employed to train the MLP network. The MLP network is implemented by the neural network toolbox in MATLAB 2017a software. For the SVR and LSSVR models, the Gaussian kernel is selected as the kernel function (Sun et al. 2018). The penalty coefficient and kernel scale of SVR and LSSVR are not cross-validated. They are set as $iqr(x)/1.349$ and 1, respectively, where $iqr(x)$ is the interquartile range of the processed target series. Regarding the VMD algorithm, the optimal mode number is set to three using the difference between the center frequencies of the adjacent subseries, as the center frequency is closely related to the decomposition results of VMD (Dragomiretskiy and Zosso 2014). The VMD algorithm is implemented using the VMD package in MATLAB 2017a software. The lag orders for tourist arrivals in machine learning models are determined using a partial mutual information method (maximum embedding order $d$=24). By this means, a short-term nonlinear dependency can be learned between the input and output data.

Using the research design mentioned above, forecasting experiments for tourist arrivals were conducted. Accordingly, the forecasting performances of all of the examined models were evaluated through the two accuracy measures.

The decomposition results of the Beijing and Hainan tourist arrival series using

VMD are shown in **Figs. 2** and **3**. We note that original passenger flow data are decomposed into trend, periodic and volatile components through the VMD algorithm. All of the periodic components of these tourist arrival series show a one-year cycle. In addition, the following measures are considered when analyzing each component, such as the mean period of each component, the correlation coefficient between the original tourist arrival series and each component, and the variance percentage of each component. **Table 4** shows the measures of each component for the Beijing and Hainan tourist arrivals. The mean period under study is defined as the value obtained by dividing the total number of points by the peak number of each component because the amplitude and frequency of a component may change continuously with time and the period is not constant. The Pearson correlation coefficient is used to measure the correlations between the original tourist arrival series and each component. However, because these components are independent of each other, it may be possible to use the variance percentage to explain the contribution of each component to the total volatility of the observed tourist arrival series. The results of two decompositions show that the dominant mode of the observed data is not the volatile part but the trend and seasonal parts. The coefficients between the original tourist arrival and periodic component reach 0.63 and 0.34 for Beijing city and Hainan province, respectively. However, the coefficients between the original tourist arrival series and trend component reach a high level of more than 0.74 and 0.93 for Beijing city and Hainan province, respectively. Additionally, the variance in the periodic component accounts for more than 31.04% of the total volatility of the observed tourist arrival data. The highest value is more than 86%. As Shen (Shen et al. 2009) noted, the periodic component is often considered the long-term behavior trend of tourist arrival flows.

**Table 4** Measures of modes for the Beijing and Hainan tourist arrivals.

| Modes | Beijing | | | Hainan | | |
|---|---|---|---|---|---|---|
| | Mean period | Correlation coefficient | Variance as % of observed | Mean period | Correlation coefficient | Variance as % of observed |
| Volatility part | 3.03 | 0.38 | 11.52 | 4.05 | 0.19 | 2.56 |
| Period part | 11.75 | 0.63 | 31.04 | 11.57 | 0.34 | 9.63 |
| Trend part | 94.00 | 0.74 | 46.71 | 162.00 | 0.93 | 85.52 |

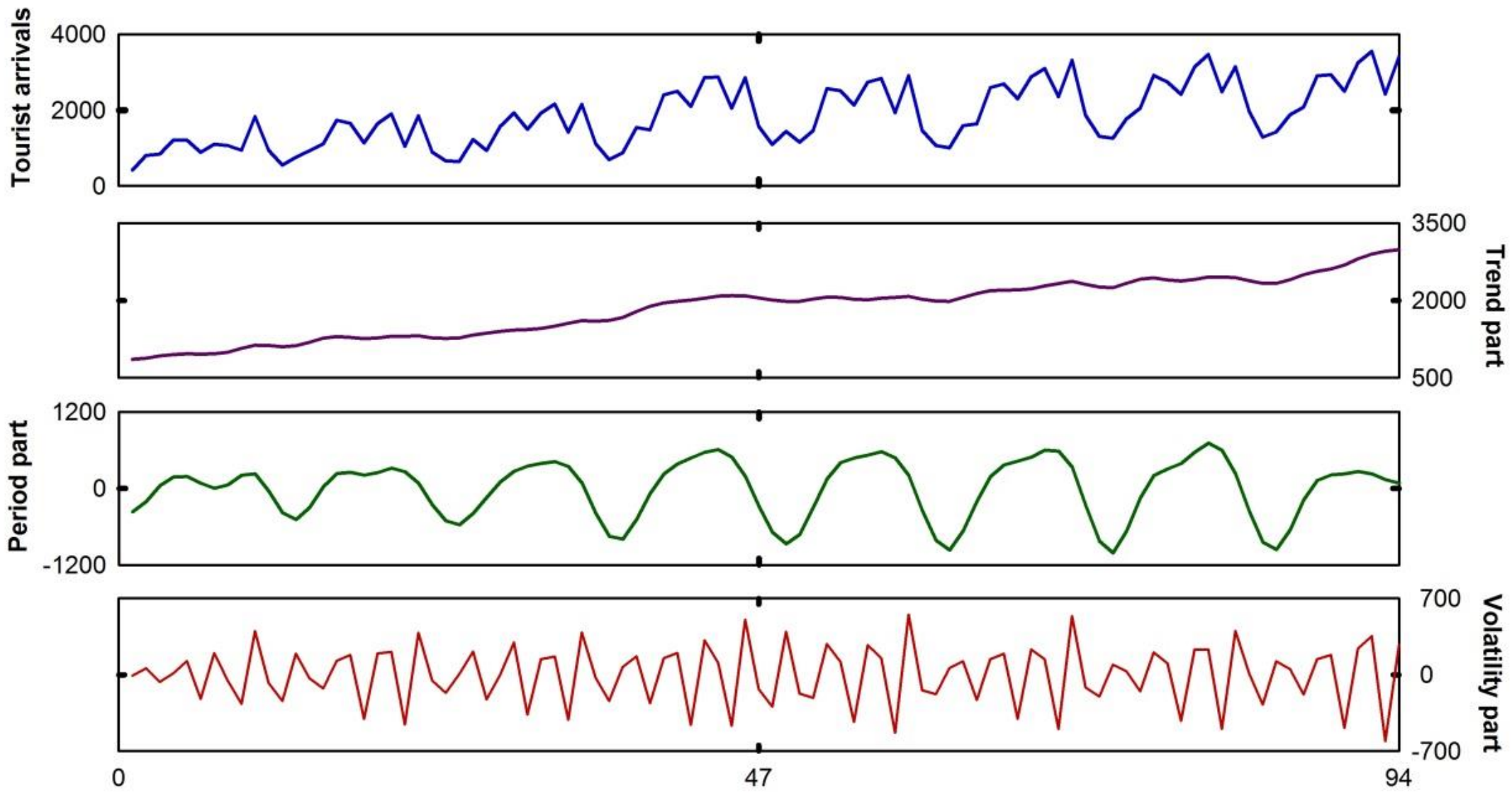

**Fig. 2** The VMD results of in-sample tourist arrivals in Beijing city.

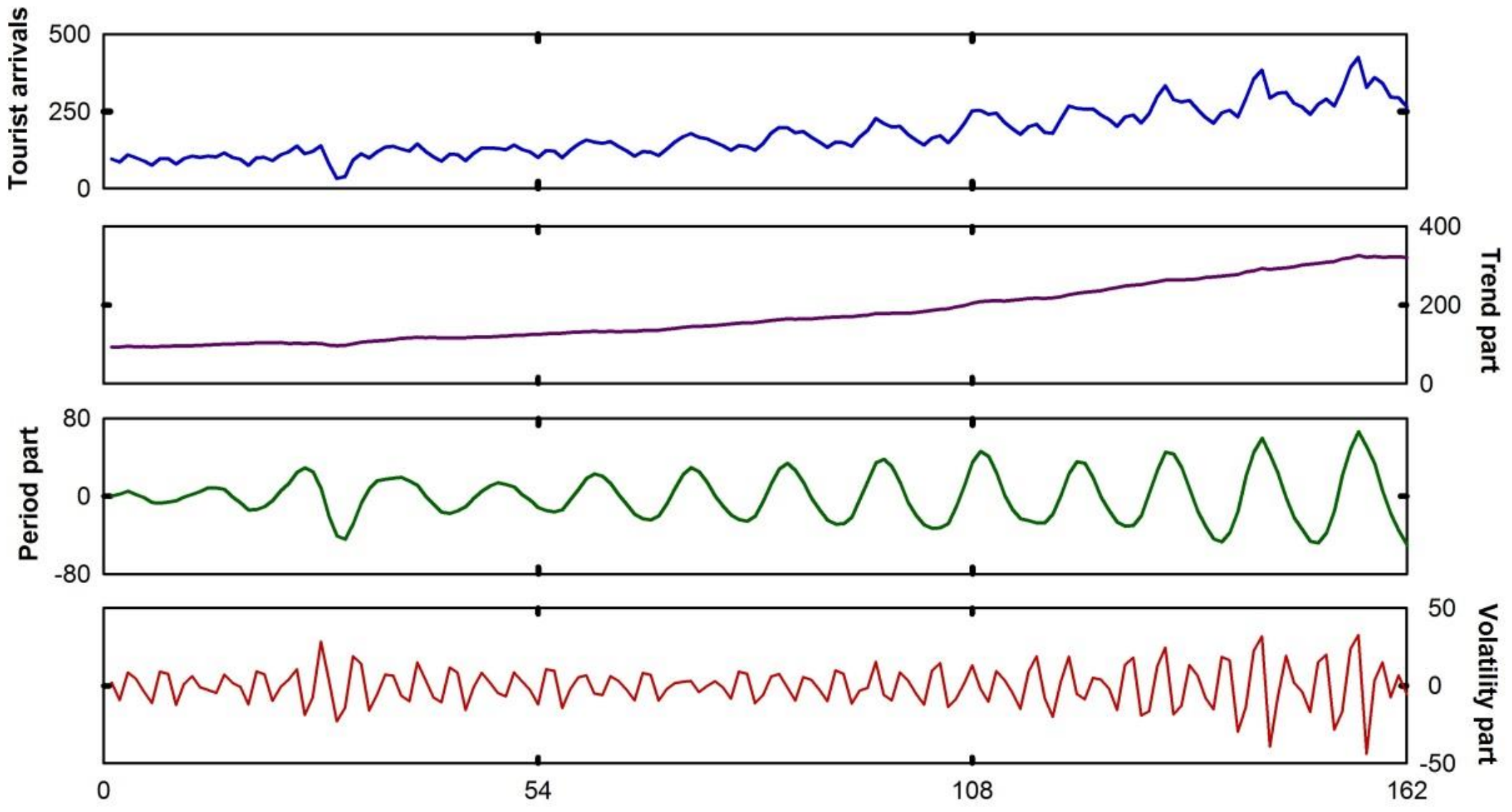

**Fig. 3** The VMD results of in-sample tourist arrivals in Hainan Province.

After the decomposition, as discussed in **Section 4**, the ARIMA model is used to forecast the extracted trend component, the SARIMA model is employed to forecast the extracted periodic component, and the LSSVR model is used to forecast the extracted volatile component. Finally, the forecasting results of the trend, periodic and volatile components are integrated into an aggregated output via another LSSVR model.

The forecasting performance of the nine models (i.e., our approach, VMD-LSSVR, VMD-SVR, VMD-MLP, LSSVR, SVR, MLP, SARIMA, and ARIMA) under study at two destinations across the three forecasting horizons ($h$-step-ahead, i.e., $h$=1,3,6) for MAPE and DS are shown in **Tables 5-7**.

**Table 5** Forecasting performance of different models; one-month-ahead forecasts.

| Strategies | Models | Beijing | | Hainan | |
|---|---|---|---|---|---|
| | | MAPE (%) | DS (%) | MAPE (%) | DS (%) |
| Single forecasts | ARIMA | 3.15 | 55.00 | 3.06 | 67.50 |
| | SARIMA | 2.14 | 80.00 | 2.39 | 80.00 |
| | MLP | 2.24 | 75.00 | 2.10 | 82.50 |
| | SVR | 2.02 | 70.00 | 1.57 | 77.50 |
| | LSSVR | 1.89 | 70.00 | 1.26 | 80.00 |
| Ensemble forecasts | VMD-MLP | 0.92 | 80.00 | 1.05 | 92.50 |
| | VMD-SVR | 0.97 | 90.00 | 0.82 | 90.00 |
| | VMD-LSSVR | 0.79 | 90.00 | 0.88 | 92.50 |
| | Our approach | 0.41 | 100.00 | 0.37 | 97.50 |

**Table 6** Forecasting performance of different models; three-month-ahead forecasts.

| Strategies | Models | Beijing | | Hainan | |
|---|---|---|---|---|---|
| | | MAPE (%) | DS (%) | MAPE (%) | DS (%) |
| Single forecasts | ARIMA | 4.25 | 50.00 | 4.37 | 62.50 |
| | SARIMA | 3.47 | 70.00 | 3.97 | 75.00 |
| | MLP | 2.62 | 70.00 | 2.22 | 77.50 |
| | SVR | 2.05 | 70.00 | 1.86 | 75.00 |
| | LSSVR | 1.92 | 65.00 | 1.16 | 75.00 |
| Ensemble forecasts | VMD-MLP | 1.06 | 75.00 | 1.03 | 87.50 |
| | VMD-SVR | 1.09 | 90.00 | 1.01 | 85.00 |
| | VMD-LSSVR | 0.90 | 85.00 | 0.82 | 87.50 |
| | Our approach | 0.63 | 95.00 | 0.58 | 95.00 |

**Table 7** Forecasting performance of different models; six-month-ahead forecasts.

| Strategies | Models | Beijing | | Hainan | |
|---|---|---|---|---|---|
| | | MAPE (%) | DS (%) | MAPE (%) | DS (%) |
| Single forecasts | ARIMA | 4.16 | 45.00 | 4.54 | 55.00 |
| | SARIMA | 3.58 | 60.00 | 3.86 | 70.00 |
| | MLP | 2.73 | 60.00 | 2.41 | 67.50 |
| | SVR | 2.11 | 65.00 | 1.97 | 70.00 |
| | LSSVR | 1.96 | 60.00 | 1.25 | 72.50 |
| Ensemble forecasts | VMD-MLP | 1.18 | 70.00 | 1.16 | 80.00 |
| | VMD-SVR | 1.14 | 85.00 | 1.07 | 77.50 |
| | VMD-LSSVR | 0.95 | 80.00 | 0.89 | 80.00 |
| | Our approach | 0.75 | 85.00 | 0.71 | 90.00 |

The results of the above tables show that our proposed adaptive multiscale ensemble learning approach is the best approach for tourist arrival forecasting among all forecasting horizons (*h*-step-ahead, i.e., *h*=1,3,6) for two destinations relative to the other eight benchmark models under study. It is conceivable that the reason behind the inferiority of the LSSVR, SVR, and MLP relative to our proposed adaptive multiscale ensemble learning approach is that there three pure artificial intelligence techniques cannot model periodic components directly, corresponding with the works of Shen et al. 2009. Therefore, prior data processing, such as time series decomposition, is critical and necessary for building a better forecaster, which is implemented as our proposed adaptive multiscale ensemble learning approach under study.

Additionally, the empirical results of all forecasting models under study show that the ARIMA and SARIMA models are consistently the worst forecasts for each tourist arrival, using the evaluation criteria and forecasting horizons. It is conceivable that the reason behind the inferiority of the ARIMA and SARIMA is that it is a class of typical linear models and cannot capture nonlinear patterns in tourist arrivals.

To further verify through statistical measures whether our proposed approach produces forecasts that are statistically significant and superior to other benchmarks, we employ the Diebold-Mariano (DM) (1995) statistic for forecasting accuracy, while the mean square error (MSE) is considered as the loss function (for more details on the test, see Diebold and Mariano, 1995). The DM statistic test is applied in the out-of-sample period of the two tourist arrivals series and three forecasting horizons. **Tables 8-10** summarize the results of the DM statistic.

**Table 8** DM test results for different models in one-month-ahead forecasts.

| Models | Our approach | VMD-LSSVR | VMD-SVR | VMD-MLP | LSSVR | SVR | MLP | SARIMA | ARIMA |
|---|---|---|---|---|---|---|---|---|---|
| | Beijing | | | | | | | | |
| Our approach | - | | | | | | | | |
| VMD-LSSVR | -1.8462 (0.0324) | - | | | | | | | |
| VMD-SVR | -2.1795 (0.0146) | -1.7985 (0.0360) | - | | | | | | |
| VMD-MLP | -2.5562 (0.0053) | -1.9123 (0.0279) | -1.6154 (0.0531) | - | | | | | |
| LSSVR | -3.8925 (0.0000) | -2.9538 (0.0016) | -3.0125 (0.0013) | -1.8941 (0.0291) | - | | | | |
| SVR | -4.0164 (0.0000) | -3.5679 (0.0002) | -3.4463 (0.0002) | -3.5129 (0.0002) | -1.7653 (0.0388) | - | | | |
| MLP | -4.2267 (0.0000) | -4.1538 (0.0000) | -4.0152 (0.0000) | -4.1146 (0.0000) | -1.9015 (0.0286) | -1.7254 (0.0422) | - | | |
| SARIMA | -5.8926 (0.0000) | -4.8766 (0.0000) | -4.7419 (0.0000) | -4.6257 (0.0000) | -1.9143 (0.0278) | -1.6568 (0.0488) | -1.6154 (0.0531) | - | |
| ARIMA | -6.4518 (0.0000) | -5.6476 (0.0000) | -6.0254 (0.0000) | -5.9168 (0.0000) | -3.6287 (0.0001) | -1.9483 (0.0257) | -1.8974 (0.0289) | -1.6154 (0.0531) | - |
| | Hainan | | | | | | | | |
| Our approach | - | | | | | | | | |
| VMD-LSSVR | -1.9896 (0.0233) | - | | | | | | | |
| VMD-SVR | -2.2567 (0.0120) | -1.8153 (0.0347) | - | | | | | | |
| VMD-MLP | -2.6159 (0.0044) | -1.9642 (0.0243) | -1.7146 (0.0432) | - | | | | | |
| LSSVR | -3.9798 (0.0000) | -3.0167 (0.0013) | -3.3314 (0.0004) | -1.9416 (0.0261) | - | | | | |
| SVR | -4.1149 (0.0000) | -3.6641 (0.0001) | -3.5143 (0.0002) | -3.6148 (0.0001) | -1.8142 (0.0348) | - | | | |
| MLP | -4.6257 (0.0000) | -4.3149 (0.0000) | -4.2349 (0.0000) | -4.2546 (0.0000) | -1.9942 (0.0231) | -1.7416 (0.0408) | - | | |
| SARIMA | -5.7846 (0.0000) | -4.9143 (0.0000) | -4.9844 (0.0000) | -4.8419 (0.0000) | -1.8946 (0.0291) | -1.6129 (0.0534) | -1.7438 (0.0406) | - | |
| ARIMA | -6.3342 (0.0000) | -6.0146 (0.0000) | -5.8941 (0.0000) | -6.0143 (0.0000) | -3.7419 (0.0001) | -1.8943 (0.0291) | -1.9062 (0.0283) | -1.6583 (0.0486) | - |

**Table 9** DM test results for different models in three-month-ahead forecasts.

| Models | Our approach | VMD-LSSVR | VMD-SVR | VMD-MLP | LSSVR | SVR | MLP | SARIMA | ARIMA |
|---|---|---|---|---|---|---|---|---|---|
| | Beijing | | | | | | | | |
| Our approach | - | | | | | | | | |
| VMD-LSSVR | -1.7891 (0.0368) | - | | | | | | | |
| VMD-SVR | -2.0269 (0.0213) | -1.8023 (0.0357) | - | | | | | | |
| VMD-MLP | -2.5782 (0.0050) | -1.9255 (0.0271) | -1.7133 (0.0433) | - | | | | | |
| LSSVR | -3.9011 (0.0000) | -3.0146 (0.0013) | -3.1426 (0.0008) | -1.9017 (0.0286) | - | | | | |
| SVR | -4.1207 (0.0000) | -3.6628 (0.0001) | -3.5369 (0.0002) | -3.4582 (0.0003) | -1.8036 (0.0356) | - | | | |
| MLP | -4.2079 (0.0000) | -4.2143 (0.0000) | -4.1596 (0.0000) | -4.2648 (0.0000) | -1.9214 (0.0273) | -1.7631(0.0389) | - | | |
| SARIMA | -5.8853 (0.0000) | -4.9017 (0.0000) | -4.6472 (0.0000) | -4.7416 (0.0000) | -1.9046 (0.0284) | -1.6244 (0.0521) | -1.6529 (0.0492) | - | |
| ARIMA | -6.3171 (0.0000) | -5.5693 (0.0000) | -6.2214 (0.0000) | -5.9861 (0.0000) | -3.7463 (0.0001) | -1.9267 (0.0270) | -1.9011 (0.0286) | -1.6354 (0.0510) | - |
| | Hainan | | | | | | | | |
| Our approach | - | | | | | | | | |
| VMD-LSSVR | -1.8859 (0.0297) | - | | | | | | | |
| VMD-SVR | -2.2368 (0.0126) | -1.8155 (0.0347) | - | | | | | | |
| VMD-MLP | -2.6129 (0.0045) | -1.9258 (0.0271) | -1.7263 (0.0421) | - | | | | | |
| LSSVR | -3.9251 (0.0000) | -3.0168 (0.0013) | -3.1042 (0.0010) | -1.9015 (0.0286) | - | | | | |
| SVR | -4.1256 (0.0000) | -3.6627 (0.0001) | -3.5833 (0.0002) | -3.6521 (0.0001) | -1.7844 (0.0372) | - | | | |
| MLP | -4.1689 (0.0000) | -4.0944 (0.0000) | -4.2105 (0.0000) | -4.2046 (0.0000) | -1.9213 (0.0273) | -1.7526 (0.0398) | - | | |
| SARIMA | -5.9014 (0.0000) | -4.9063 (0.0000) | -4.8124 (0.0000) | -4.5632 (0.0000) | -1.9014 (0.0286) | -1.6935 (0.0452) | -1.6354 (0.0510) | - | |
| ARIMA | -6.5528 (0.0000) | -5.6548 (0.0000) | -6.1146 (0.0000) | -5.9016 (0.0000) | -3.7329 (0.0001) | -1.9844 (0.0236) | -1.9215 (0.0273) | -1.5638 (0.0589) | - |

**Table 10** DM test results for different models in six-month-ahead forecasts.

| Models | Our approach | VMD-LSSVR | VMD-SVR | VMD-MLP | LSSVR | SVR | MLP | SARIMA | ARIMA |
|---|---|---|---|---|---|---|---|---|---|
| | Beijing | | | | | | | | |
| Our approach | - | | | | | | | | |
| VMD-LSSVR | -1.8926 (0.0292) | - | | | | | | | |
| VMD-SVR | -2.3567 (0.0092) | -1.8137 (0.0349) | - | | | | | | |
| VMD-MLP | -2.4692 (0.0068) | -1.9341 (0.0266) | -1.7016 (0.0444) | - | | | | | |
| LSSVR | -3.9628 (0.0000) | -2.9046 (0.0018) | -2.9972 (0.0014) | -1.8526 (0.0320) | - | | | | |
| SVR | -4.1563 (0.0000) | -3.4168 (0.0003) | -3.3105 (0.0005) | -3.7036 (0.0001) | -1.8016 (0.0358) | - | | | |
| MLP | -4.3367 (0.0000) | -4.0143 (0.0000) | -4.1046 (0.0000) | -4.0416 (0.0000) | -1.9246 (0.0271) | -1.7361 (0.0413) | - | | |
| SARIMA | -5.9027 (0.0000) | -4.9168 (0.0000) | -4.8122 (0.0000) | -4.7019 (0.0000) | -1.9517 (0.0255) | -1.7018 (0.0444) | -1.6528 (0.0492) | - | |
| ARIMA | -6.5149 (0.0000) | -5.7458 (0.0000) | -6.1526 (0.0000) | -5.8743 (0.0000) | -3.7469 (0.0001) | -1.9910 (0.0232) | -1.9014 (0.0286) | -1.6842 (0.0461) | - |
| | Hainan | | | | | | | | |
| Our approach | - | | | | | | | | |
| VMD-LSSVR | -1.8793 (0.0301) | - | | | | | | | |
| VMD-SVR | -2.3268 (0.0100) | -1.8974 (0.0289) | - | | | | | | |
| VMD-MLP | -2.7816 (0.0027) | -1.9568 (0.0252) | -1.7956 (0.0363) | - | | | | | |
| LSSVR | -4.0168 (0.0000) | -3.1462 (0.0008) | -3.2152 (0.0007) | -1.9365 (0.0264) | - | | | | |
| SVR | -4.2249 (0.0000) | -3.7063 (0.0001) | -3.6085 (0.0002) | -3.5478 (0.0002) | -1.7952 (0.0363) | - | | | |
| MLP | -4.2761 (0.0000) | -4.2571 (0.0000) | -4.1763 (0.0000) | -4.1462 (0.0000) | -1.9657 (0.0247) | -1.7843 (0.0372) | - | | |
| SARIMA | -5.9683 (0.0000) | -4.8914 (0.0000) | -4.9143 (0.0000) | -4.6185 (0.0000) | -1.9851 (0.0236) | -1.7916 (0.0366) | -1.6987 (0.0447) | - | |
| ARIMA | -6.4472 (0.0000) | -5.7416 (0.0000) | -6.2108 (0.0000) | -6.1258 (0.0000) | -3.8862 (0.0001) | -2.0168 (0.0219) | -1.9681 (0.0245) | -1.6049 (0.0543) | - |

From the above three tables, we note that the null hypothesis of equal forecasting accuracy is rejected for all comparisons and loss functions at the 5% confidence interval since all the absolute values of test statistics are higher than the critical value of 1.69. Moreover, the statistical superiority of our proposed approach forecasts is confirmed, as the realizations in the DM statistic are negative for both loss functions.

In addition, the Pesaran-Timmermann (PT) statistic test is employed to examine whether the directional changes in the actual and predicted values are the same. In other words, it determines how well increases and decreases in the predicted value follow real increases and decreases of the tourist arrival time series. The null hypothesis is that the model under study has no power in forecasting the relevant tourist arrivals. The out-of-sample statistical performance of all models for the Beijing city and Hainan province tourist arrivals is provided in **Table 11**.

The results in **Table 11** show that the PT statistic tests reject the null hypothesis of no forecasting power at the 1% confidence interval for all multiscale ensemble approaches and the series under study. In other words, it indicates that all multiscale ensemble approaches are capable of capturing the directional movements of tourist arrivals. Based on the above results, our proposed adaptive multiscale ensemble learning approach is the superior model utilizing statistical efficiency.

**Table 11** PT test results for different models.

| Models | Beijing | | | Hainan | | |
|---|---|---|---|---|---|---|
| | 1-month-ahead | 3-month-ahead | 6-month-ahead | 1-month-ahead | 3-month-ahead | 6-month-ahead |
| Our approach | 4.5129 (0.0000) | 4.2068 (0.0000) | 3.3419 (0.0008) | 4.6672 (0.0000) | 4.3284 (0.0000) | 3.5697 (0.0004) |
| VMD-LSSVR | 3.8512 (0.0001) | 3.7743 (0.0002) | 2.8146 (0.0049) | 3.9253 (0.0001) | 3.8142 (0.0001) | 3.0164 (0.0026) |
| VMD-SVR | 3.2954 (0.0010) | 3.3018 (0.0010) | 2.4413 (0.0146) | 3.5826 (0.0003) | 3.4069 (0.0007) | 2.6418 (0.0082) |
| VMD-MLP | 3.2352 (0.0012) | 3.0139 (0.0026) | 2.1743 (0.0297) | 3.4367 (0.0006) | 3.1587 (0.0016) | 2.5267 (0.0115) |
| LSSVR | 2.0215 (0.0432) | 2.1127 (0.0346) | 1.9162 (0.0553) | 2.3148 (0.0206) | 2.2248 (0.0261) | 1.9456 (0.0517) |
| SVR | 1.9625 (0.0497) | 1.8954 (0.0580) | 1.8514 (0.0641) | 1.9164 (0.0553) | 1.8834 (0.0596) | 1.8016 (0.0716) |
| MLP | 1.7489 (0.0803) | 1.7016 (0.0888) | 1.6082 (0.1078) | 1.8107 (0.0702) | 1.7821 (0.0747) | 1.6149 (0.1063) |
| SARIMA | 1.6541 (0.0981) | 1.5913 (0.1115) | 1.5316 (0.1256) | 1.6638 (0.0962) | 1.6025 (0.1090) | 1.5824 (0.1136) |
| ARIMA | 1.5892 (0.1120) | 1.4018 (0.1610) | 1.4139 (0.1574) | 1.5329 (0.1253) | 1.5036 (0.1327) | 1.4673 (0.1423) |

### *5.3 Discussions*

Overall, from the above analysis of the empirical results obtained in this study,

some interesting findings can be drawn as follows. (1) LSSVR performs better than all other single benchmark models. (2) Through the comparison between the VMD-based multiscale ensemble approach and their corresponding single model, the VMD-based multiscale ensemble approach is the winner. This means that mode decomposition of the tourist arrival time series before further forecasting can effectively improve the forecasting performance for tourist arrival forecasting. (3) Due to the highly nonlinear and periodic patterns in the tourist arrival series, AI-based nonlinear models are more suitable for forecasting time series with periodic volatility than linear models. (4) Our proposed adaptive multiscale ensemble learning approach is consistently the best approach relative to all other benchmark models understudied for tourist arrival forecasting utilizing statistical accuracy and forecasting horizons. (5) Our proposed adaptive multiscale ensemble learning approach can be regarded as a promising framework for forecasting time series with highly periodic volatility.

## 6 Conclusions

The purpose of this study was to accurately forecast seasonal and trend tourism arrivals. The main idea of our new adaptive multiscale ensemble (AME) learning approach is as follows: (1) First, we apply the VMD decomposition method to divide the original tourist arrival time series into trend, periodic and volatility components. (2) Each mode series is forecasted by using ARIMA, SARIMA, and LSSVR. (3) The forecasting results of all extracted modes are fused to generate a final forecasting result using LSSVR as an ensemble learning method. The experimental results suggest that the VMD-based multiscale ensemble approach has the best performance, and our proposed adaptive multiscale ensemble learning approach is consistently the best relative to all other benchmark models understudied for tourist arrival forecasting utilizing statistical accuracy and forecasting horizons.

Our study can provide some managerial insights. First, from the perspective of practitioners, it is very important to determine whether the model predicts the change in direction as expected. Our proposed adaptive multiscale ensemble learning approach can be regarded as a promising framework for forecasting time series with periodic volatility. In other words, the results of the Pesaran-Timmermann (PT) statistic test tell us that our proposed model can assist practitioners in making correct decisions to respond to the need for emergency management of tourism safety. After the PT statistic test, it shows that our proposed adaptive multiscale ensemble learning approach is the superior model using statistical efficiency. Second, tourism forecasting, especially tourist demand, is of great significance to national tourism security emergency management. The time series of tourism contains information on tourist behaviors, and its periodicity and seasonality have a great influence on forecasting models. Therefore, the decomposition work based on the original time series data proposed in this paper fundamentally determines the correct choice of subsequent models, and the forecasting accuracy of the combination forecast is very superior.

In addition to tourist arrival forecasting, our proposed adaptive multiscale ensemble learning approach can be regarded as a promising framework for forecasting time series with highly periodic volatility, including stock trend forecasting, crude oil price forecasting, and exchange rate forecasting.

However, this study has some limitations since it only focuses on univariate time series analysis without considering other factors affecting tourist arrivals. If those factors are integrated into our proposed approach, the forecasting performance may be improved. In addition, we use LSSVR, a nonlinear AI model, to integrate the forecasting results. According to the study of Shen (Shen et al. 2008), the performance of ensemble learning is affected by the features of each base learner. Our future research intends to select different integrated learners to improve the forecasting accuracy of the combined forecasts.

## 7 Conflict of interests

The authors declare that there are no conflicts of interest regarding the publication of this paper.

## 8 Acknowledgment

This research work was partly supported by the National Natural Science Foundation of China under Grants No. 71801213, No. 71771208 and No. 71642006 and a grant from the Research Grants Council of the Hong Kong Special Administrative Region, China (Project No. T32-101/15-R).

# References:

Andrawis, R.R., A.F. Atiya, and H. El-Shishiny. 2011. "Combination of long term and short term forecasts, with application to tourism demand forecasting." International Journal of Forecasting 27:870-886.

Athanasopoulos, G., H. Song, and J.A. Sun. 2017. "Bagging in Tourism Demand Modeling and Forecasting." Journal of Travel Research 57:52-68.

Athanasopoulos, G., R.J. Hyndman, H. Song, and D.C. Wu. 2011. "The tourism forecasting competition." International Journal of Forecasting 27:822-844.

Athanasopoulos, G., and R.J. Hyndman. 2008. "Modelling and forecasting Australian domestic tourism." Tourism Management 29:19-31.

Bangwayo-Skeete, P.F., and R.W. Skeete. 2015. "Can Google data improve the forecasting performance of tourist arrivals? Mixed-data sampling approach." Tourism Management 46:454-464.

Box, G. E., Jenkins, G. M., Reinsel, G. C., & Ljung, G. M. 2015. "*Time series analysis: forecasting and control*". John Wiley & Sons.

Cang, S. 2014. "A Comparative Analysis of Three Types of Tourism Demand Forecasting Models: Individual, Linear Combination and Non-linear Combination." International Journal of Tourism Research 16:596-607.

Cao, Z., G. Li, and H. Song. 2017. "Modelling the interdependence of tourism demand: The global vector autoregressive approach." Annals of Tourism Research 67:1-13.

Chan, C.K., S.F. Witt, Y.C.E. Lee, and H. Song. 2010. "Tourism forecast combination using the CUSUM technique." Tourism Management 31:891-897.

Chen, C., M. Lai, and C. Yeh.2012. "Forecasting tourism demand based on empirical mode decomposition and neural network." Knowledge-Based Systems 26:281-287.

Chen, J.L., G. Li, D.C. Wu, and S. Shen.2017. "Forecasting Seasonal Tourism Demand Using a Multiseries Structural Time Series Method." Journal of Travel Research 58:92-103.

Chen, K., and C. Wang. 2007. "Support vector regression with genetic algorithms in forecasting tourism demand." Tourism Management 28:215-226.

Chu, F. 2004. "Forecasting tourism demand: a cubic polynomial approach." Tourism Management

25:209-218.

Chu, F. 2008. "Analyzing and forecasting tourism demand with ARAR algorithm." Tourism Management 29:1185-1196.

Chu, F. 2009. "Forecasting tourism demand with ARMA-based methods." Tourism Management 30:740-751.

Chu, F. 2011. "A piecewise linear approach to modeling and forecasting demand for Macau tourism." Tourism Management 32:1414-1420.

Coshall, J.T. 2009. "Combining volatility and smoothing forecasts of UK demand for international tourism." Tourism Management 30:495-511.

Coshall, J.T., and R. Charlesworth. 2011. "A management orientated approach to combination forecasting of tourism demand." Tourism Management 32:759-769.

Cortes, C., & Vapnik, V. 1995. "Support-vector networks." Machine learning, 20(3), 273-297.

Dragomiretskiy, K., and D. Zosso. 2014. "Variational Mode Decomposition." IEEE Transactions on Signal Processing 62:531-544.

du Preez, J., and S.F. Witt. 2003. "Univariate versus multivariate time series forecasting: an application to international tourism demand." International Journal of Forecasting 19:435-451.

Fildes, R., Y. Wei, and S. Ismail. 2011. "Evaluating the forecasting performance of econometric models of air passenger traffic flows using multiple error measures." International Journal of Forecasting 27:902-922.

Gil-Alana, L.A., J. Cunado, and F. Perez De Gracia. 2008. "Tourism in the Canary Islands: forecasting using several seasonal time series models." Journal of Forecasting 27:621-636.

Godarzi, A.A., R.M. Amiri, A. Talaei, and T. Jamasb. 2014. "Predicting oil price movements: A dynamic Artificial Neural Network approach." Energy Policy 68:371-382.

Goh, C., and R. Law. 2003. "Incorporating the rough sets theory into travel demand analysis." Tourism Management 24:511-517.

Guizzardi, A., and M. Mazzocchi. 2010. "Tourism demand for Italy and the business cycle." Tourism Management 31:367-377.

Gunter, U., and I. Önder. 2015. "Forecasting international city tourism demand for Paris: Accuracy of uni- and multivariate models employing monthly data." Tourism Management 46:123-135.

Gunter, U., and I. Önder. 2016. "Forecasting city arrivals with Google Analytics." Annals of Tourism

Research 61:199-212.

Hassani, H., E.S. Silva, N. Antonakakis, G. Filis, and R. Gupta. 2017. "Forecasting accuracy evaluation of tourist arrivals." Annals of Tourism Research 63:112-127.

Hirashima, A., J. Jones, C.S. Bonham, and P. Fuleky. 2017. "Forecasting in a Mixed Up World: Nowcasting Hawaii Tourism." Annals of Tourism Research 63:191-202.

Kim, J.H., H. Song, and K.K.F. Wong. 2010. "Bias-Corrected Bootstrap Prediction Intervals for Autoregressive Model: New Alternatives with Applications to Tourism Forecasting." Journal of Forecasting 29:655-672.

Koenig-Lewis, N., and E.E. Bischoff. 2005. "Seasonality research: the state of the art." International Journal of Tourism Research 7:201-219.

Li, S., T. Chen, L. Wang, and C. Ming. 2018. "Effective tourist volume forecasting supported by PCA and improved BPNN using Baidu index." Tourism Management 68:116-126.

Li, X., B. Pan, R. Law, and X. Huang. 2017. "Forecasting tourism demand with composite search index." Tourism Management 59:57-66.

Liu, Y., F. Tseng, and Y. Tseng. 2018. "Big Data analytics for forecasting tourism destination arrivals with the applied Vector Autoregression model." Technological Forecasting and Social Change 130:123-134.

Olmedo, E. 2016. "Comparison of Near Neighbour and Neural Network in Travel Forecasting." Journal of Forecasting 35:217-223.

Palmer, A., J. José Montaño, and A. Sesé. 2006. "Designing an artificial neural network for forecasting tourism time series." Tourism Management 27:781-790.

Pan, B., D. Chenguang Wu, and H. Song. 2012. "Forecasting hotel room demand using search engine data." Journal of Hospitality and Tourism Technology 3:196-210.

Pan, B., and Y. Yang. 2016. "Forecasting Destination Weekly Hotel Occupancy with Big Data." Journal of Travel Research 56:957-970.

Peng, B., H. Song, and G.I. Crouch. 2014. "A meta-analysis of international tourism demand forecasting and implications for practice." Tourism Management 45:181-193.

Rivera, R. 2016. "A dynamic linear model to forecast hotel registrations in Puerto Rico using Google Trends data." Tourism Management 57:12-20.

Shahrabi, J., E. Hadavandi, and S. Asadi. 2013. "Developing a hybrid intelligent model for forecasting

problems: Case study of tourism demand time series." Knowledge-Based Systems 43:112-122.

Shen, S., G. Li, and H. Song. 2008. "An Assessment of Combining Tourism Demand Forecasts over Different Time Horizons." Journal of Travel Research 47:197-207.

Shen, S., G. Li, and H. Song. 2009. "Effect of Seasonality Treatment on the Forecasting Performance of Tourism Demand Models." Tourism Economics 15:693-708.

Shen, S., G. Li, and H. Song. 2011. "Combination forecasts of International tourism demand." Annals of Tourism Research 38:72-89.

Song, H., B.Z. Gao, and V.S. Lin. 2013. "Combining statistical and judgmental forecasts via a web-based tourism demand forecasting system." International Journal of Forecasting 29:295-310.

Song, H., G. Li, S.F. Witt, and B. Fei. 2010. "Tourism Demand Modelling and Forecasting: How Should Demand Be Measured?" Tourism Economics 16:63-81.

Song, H., G. Li, S.F. Witt, and G. Athanasopoulos. 2011. "Forecasting tourist arrivals using time-varying parameter structural time series models." International Journal of Forecasting 27:855-869.

Song, H., K.K.F. Wong, and K.K.S. Chon. 2003. "Modelling and forecasting the demand for Hong Kong tourism." International Journal of Hospitality Management 22:435-451.

Song, H., S.F. Witt, and T.C. Jensen. 2003. "Tourism forecasting: accuracy of alternative econometric models." International Journal of Forecasting 19:123-141.

Song, H., and S.F. Witt. 2006. "Forecasting international tourist flows to Macau." Tourism Management 27:214-224.

Sun, S., Y. Wei, K. Tsui, and S. Wang. 2019. "Forecasting tourist arrivals with machine learning and internet search index." Tourism Management 70:1-10.

Sun, X., W. Sun, J. Wang, Y. Zhang, and Y. Gao. 2016. "Using a Grey–Markov model optimized by Cuckoo search algorithm to forecast the annual foreign tourist arrivals to China." Tourism Management 52:369-379.

Suykens, J. A., & Vandewalle, J. 1999. "Chaos control using least‐squares support vector machines." International journal of circuit theory and applications 27(6):605-615.

Wang, C. 2004. "Predicting tourism demand using fuzzy time series and hybrid grey theory." Tourism Management 25:367-374.

Wong, K.K.F., H. Song, S.F. Witt, and D.C. Wu. 2007. "Tourism forecasting: To combine or not to combine?" Tourism Management 28:1068-1078.

Wong, K.K.F., H. Song, and K.S. Chon. 2006. "Bayesian models for tourism demand forecasting." Tourism Management 27:773-780.

Wu, L., and G. Cao. 2016. "Seasonal SVR with FOA algorithm for single-step and multi-step ahead forecasting in monthly inbound tourist flow." Knowledge-Based Systems 110:157-166.

Wu, Q., R. Law, and X. Xu. 2012. "A sparse Gaussian process regression model for tourism demand forecasting in Hong Kong." Expert Systems with Applications 39:4769-4774.

Yang, X., B. Pan, J.A. Evans, and B. Lv. 2015. "Forecasting Chinese tourist volume with search engine data." Tourism Management 46:386-397.